\documentclass[12pt,letterpaper]{article}

\usepackage{amsfonts}
\usepackage{graphicx}
\usepackage{amssymb}
\usepackage{amsmath}
\usepackage[T1]{fontenc}
\usepackage{mathdots}
\usepackage{mathrsfs}
\newcommand{\Lie}{{\mathscr{L}}}

\newcommand{\R}{{\mathbb R}}

\newcommand{\Z}{{\mathbb Z}}

\newcommand{\LL}{{\mathcal L}}

\newcommand{\Spin}{{\mathbb S}}

\newcommand{\mA}{{m_{\rm R{\overline L}}}}
\newcommand{\mB}{{m_{\rm L{\overline R}}}}

\title{\Large Spinor Lie derivatives\\ and\\ Fermion 
stress--energies}

\author{\large A. D. Helfer\\%
Department of Mathematics, and \\%
Department of Physics \& Astronomy\\%
University of Missouri\\%
Columbia, MO 65211, U.S.A.%
\thanks{helfera@missouri.edu}}

\begin{document}

\maketitle




\begin{abstract}
Stress--energies for Fermi fields are derived from the principle of general covariance.  This is done by developing a notion of Lie derivatives of spinors along arbitrary vector fields.  A substantial theory of such derivatives was first introduced by Kosmann; here I show how an apparent conflict in the literature on this is due to a difference in the definitions of spinors, and that tracking the Lie derivative of the Infeld--van der Waerden symbol, as well as the spinor fields under consideration, gives a fuller picture of the geometry and leads to the Fermion stress--energy.  The differences in the  definitions of spinors do not affect the results here, but could matter in certain quantum-gravity programs and for spinor transformations under discrete symmetries.
\end{abstract}

\centerline{Keywords:  Fermions, stress--energy, Lie derivatives of spinors, discrete symmetries}

\section{Introduction}

One of the beautiful and deep features of general relativity is that the principle of covariance gives rise to locally conserved source terms, that is, stress--energy tensors, for classical matter governed by Lagrangians.  For such matter, one has a universal prescription for finding the sources --- one varies the action with respect to the metric --- and this universality can be viewed
as a strengthened form of the weak equivalence principle, showing that matter fields with identical Lagrangians give identical gravitational contributions.

For relativistic quantum fields, the stress--energy becomes an operator, and so cannot directly be a source for the classical gravitational field; the question of how quantum systems give rise to gravitational fields is a major unsolved problem.  But it seems inescapable that the stress--energy will be critical in this.

\maketitle

The stress--energies of quantum field theories are only partially understood.  Part of this is due to their operator-theoretic character, and associated issues of regularization and renormalization \cite{Wald1994,ADH1996}, but these will 
not be considered here.  Rather, I want to look at the more primitive question of how the formal expression for the stress--energy is derived and justified, before one takes up the quantization.  

At this level, one can simply treat Bose fields as classical quantities.  However, for Fermi fields, there are two difficulties.  One is that, even if we set aside their operator character (in the usual sense), we must still consider them as Grassmann, rather than classical quantities.  (This issue can be dodged to some degree for purely Dirac fields, but must be faced when Majorana mass terms are allowed.)  
To understand the second, recall that what one needs for the usual general-relativistic derivation of the stress--energy is that the effect of an infinitesimal diffeomorphism on the action can be split into a variation of the matter fields and one of the metric.  However, 
because spinors depend, for their very definition, on the metric structure of space--time, one has no such split.
There is no natural action of the diffeomorphisms on spinors independent of the metric; complementarily, there is no way of varying the metric without affecting the spinors.
The conventional
general-relativistic approach to the stress--energy appears stymied.

There are, of course, known ways of circumventing these difficulties.  Indeed, the tradition in quantum field theory (as opposed to general relativity) is to determine stress--energies by different arguments (e.g. \cite{Weinberg1995v1}).  One starts with a `canonical' stress--energy (based on Hamiltonian theory in Minkowski space), then `improves' this using global special-relativistic arguments.\footnote{The canonical stress--energy $T_{ab}^{\rm can}$ is locally conserved but not symmetric, and the usual derivation of it uses strongly the translational invariance of Minkowski space; that derivation justifies its use to form the total energy--momentum as an integral $P_a=\int T_{ab}^{\rm can}\,d\Sigma ^b$ over a suitable spacelike hypersurface $\Sigma$ (and as such is adequate for many applications), but does not really explicate its local kinematic content.  To obtain a symmetric conserved stress--energy, one can add certain `correction' terms, found using global Lorentz invariance; the result is the {\em Belinfante tensor} $T_{ab}^{\rm Bel}$, which will also produce the correct total angular momentum.}
If one wants a result valid for curved space--time one must then look for an extension of the formula one has found, and verify that it applies in such cases.\footnote{Rosenfeld showed that, for tensor theories, the Belinfante tensor, generalized to curved space--time (I will call this the {\em Belinfante--Rosenfeld tensor}), will coincide with the general-relativistic stress--energy \cite{Rosenfeld1940}.}  Another route is to retain a variational approach but introduce auxiliary fields (usually a tetrad) \cite{HHKN}.\footnote{In the case of spinors, the Belinfante--Rosenfeld tensor can be obtained by varying a tetrad \cite{Rosenfeld1940}; by itself, this argument does not show that this tensor can be derived from covariance in the same way those for tensor theories are.}  Finally, one can simply guess the stress--energy, and then verify it has the desired properties.  However, none of these methods approaches the simplicity, generality, and physical depth of the classical argument.

One is thus left trying to understand in just what sense Fermion contributions to stress--energy are different from Bose ones.  Certainly Fermi fields do not have direct c-number classical counterparts, as do Bose fields.  But beyond that, as things stand, it has seemed that Fermi stress--energies could not be simply derived from local covariance, in contrast to Bose ones.  If this were the case, it would seem to hint that we should look for some other, presumably foundational, principle to determine them.  Indeed, this is one way of viewing proposals like the Kibble--Sciama theory \cite{Kibble1961,Sciama1964},\footnote{This is often referred to as the Einstein--Cartan--Kibble--Sciama theory, but as spin was unknown at the times of Einstein's and Cartan's contributions, it is here better associated with Kibble and Sciama.} as attempts to find alternate variational principles incorporating Fermions in the field equations.

In this paper, I will show that Fermion stress--energies can indeed be derived by applying the principle of general covariance to their conventional Lagrangians, without introducing auxiliary fields.  Although this cannot be done by varying the metric independently of the spinors, the covariance can be taken into account by developing the spinor geometry, and its response to diffeomorphisms, more fully than has been done.  
This can be viewed as developing a notion of Lie differentiation of spinors along arbitrary vector fields.

Because spinors are not merely differential-topological objects but differ\-en\-tial-geometric ones (that is, they depend on the metric and not just the manifold), the ordinary theory of flows and Lie derivatives does not apply to them.  (Flowing along a vector field generally distorts the metric and thus does not preserve the definition of spinors.)  This means that there is no clear set of criteria for what the Lie derivative of a spinor should be.


A definition of Lie differentiation of spinors along arbitrary vector fields was first proposed by Kosmann \cite{Kosmann1971}.  It was based on using the metric to project the action of the Lie derivative to the Clifford algebra.  This use of the metric is artificial from the point of view of ordinary Lie theory, and that has led to much work discussing ways in which it might be justified (usually by embedding that theory into a more general one; see, for example, \cite{BourguignonGauduchon1992,GodinaMatteucci2003,FatibeneFrancaviglia2003,LRW2014}, and references therein).  Kosmann's approach has been extensively used in mathematics and in somewhat formal physical arguments, but has received almost no attention in relativity.

Relativists have a long tradition of viewing spinors, not just as geometric objects built on top of a space--time, but as coding geometry within the space--time itself; this is Penrose's notion of `null flags' \cite{PR1984}.  
(One could say that spinors are not merely `vertical' objects, that is, elements of certain fibre bundles, but have `horizontal' significance, within the space--time, as well.)
It is clear that no definition of Lie differentiation of spinors along arbitrary vector fields can be developed which will preserve this:  if the vector field is not conformally Killing, it will not preserve nullness of vectors, and so it cannot preserve the geometric interpretation of spinors.  Thus the geometric significance of Kosmann's Lie derivative seemed limited.  Indeed, Kosmann's definition has spinors responding, infinitesimally, only to the rotational (or, in the space--time case, Lorentz) part of the motion induced by the flow, which seems to mean that most of the potential geometric content (the shear and divergence parts) has somehow been discarded; this sharply contrasts with what happens for tensors.
The matter is further obscured by the fact that the formula of Penrose and Rindler \cite{PR1986} for the Lie derivative along a conformal Killing field does not agree with Kosmann's (restricted to that case); Godina and Matteucci \cite{GodinaMatteucci} suggested that Penrose and Rindler worked not with spinors but with conformal spinors.

I will show here that the apparent conflicts are due to slightly different definitions of spinors used by the different authors.  These definitions are equivalent for any given metric, but the isomorphisms between them depend on the metric's scale; this gives them distinct conformal behaviors and leads to distinct Lie derivatives.  (So Penrose and Rindler's definition is self-consistent and does indeed work with spinor fields, but their spinor fields and those of Kosmann and followers are relatively conformally weighted.)  A parameter $w$ interpolating between the definitions will be introduced, and computations made for any choice of the parameter.  I will actually argue that another choice, neither Kosmann's nor that of Penrose and Rindler, gives the most natural combination of analytic and geometric properties:  the spinors will be identified with elements of the covering space of the anti-self-dual two-planes (with a `weight' $w$).  This geometric characterization will be the basis for the analysis.

A more directly geometric construction of the Lie derivatives will be given.  It is based on constructing an extension of flows which can be used to pull back spinors; then Lie derivatives are formed by differentiating these pull-backs, as usual.  It is easily seen to be equivalent to Kosmann's (allowing for the choice in how spinors are defined; Kosmann in fact used an integrated form of her Lie derivatives to define a map which is essentially the extended notion of flow used here), but, because the spinors are characterized geometrically and the generalized flows are defined first the geometric significance is emphasized.  

It will be apparent from the present perspective that one cannot really regard the Lie derivative of an individual spinor field as coding all the relevant information of the flow's action on the field's geometry.
One must at the same time keep track of the derivatives of the Infeld--van der Waerden symbol $\sigma _a{}^{AA'}$ and the alternating spinor $\epsilon _{AB}$, because these are the basis for the tensor-spinor correspondence and therefore for the geometric interpretation of spinors.  Kosmann computed equivalent quantities (the Lie derivatives of the Dirac gammas); really, what I do here is point out that these must be regarded as accompanying the Lie derivative of the spinor.

It is the derivatives of the Infeld--van der Waerden symbol which code how flow along a general vector field distorts the tensor-spinor correspondence, and which (therefore) are sensitive to the shear and divergence of the flow.  In the application to variations of Fermion Lagrangians, we will find that this distortion is in fact precisely the source of the stress--energy.  (Other contributions which might in principle have arisen, for instance curvature terms from variations of the covariant derivatives of the spinors with respect to the metric in curved space--time, cancel out.)  

The stress--energy of a standard Fermion Lagrangian with both Dirac and Majorana mass terms will be computed.  The result can be put in the same form as the conventional `symmetrized' (Belinfante--Rosenfeld) stress--energy in the Dirac case, with no mass terms appearing explicitly,\footnote{The weight $w$ chosen for the definition of spinor fields does not affect the answer, since non-trivial net contributions of $w$-dependent terms can only occur for spinor-valued functions, and the stress--energy is tensor-valued.} but the calculations (particularly in curved space--time) are not at all trivial.  This result is not surprising, since it is the only natural quadratic candidate with the requisite properties.  What we learn about the stress--energy is not so much the formula, as that that formula is a consequence of general covariance in much the same pattern as are classical stress--energies.

The plan of the paper is this.  Section 2 reviews the geometry of two-spinors, and also their connection with four-spinors.  The Lie derivatives of spinor fields are defined and discussed in Section 3; in Section 4, the stress--energy of Fermion Lagrangians is computed.  Section 5 recapitulates the main results and contains some further discussion.

{\em Notation and conventions.}  Although the general procedure here could be adapted to other dimensions and signatures, it will be given here in the physical case of four-dimensions only.  The calculus of two-spinors makes many of the computations easier and also is tied more directly to the geometric interpretations involved than would a four-spinor one.

The two-spinor conventions are those of Penrose and Rindler \cite{PR1984,PR1986} and are standard in general relativity, while the four-spinor ones are those of Schweber \cite{Schweber1961} and are compatible with common ones in quantum field theory; these books will also serve as references for most matters not fully explained here; see also Cheng and Li \cite{ChengLi} for a brief treatment of Majorana Fermions.  
The two-spinor--four-spinor correspondence we use is compatible with that common in quantum field theory.\footnote{It is different from the correspondence used by Penrose and Rindler, which is based on  an opposite sign for the Clifford algebra to what is generally found in quantum field theory.  However, this difference will be of no significance for us, since Penrose and Rindler do all of their space--time geometry with two-spinors and only mention the correspondence with four-spinors for completeness.}  With these conventions, we have an exact correspondence between the standard basis on pp. 120--125 of Penrose and Rindler \cite{PR1984} or pp. 6--8 of \cite{PR1986} and the Weyl basis on p. 79 of Schweber \cite{Schweber1961}.

Penrose's abstract-index convention will be used:  tensor and spinor indices do not take numerical values, but are used to represent the kind of object under consideration and the invariant operations of contraction, symmetrization and skew-symmetrization.
For this reason, the Infeld--van der Waerden symbol is for us a single tensor--spinor quantity, not a collection of components, and likewise for the Dirac symbol $\gamma _a$.
It will sometimes be useful, particularly when several indices are involved, to indicate certain contractions between adjacent symbols with a centered dot:  thus $\xi\cdot\nabla X^{aB} =\xi ^p\nabla _q X^{aB}$.

The metric has signature $+{}-{}-{}-$, and the curvature satisfies $[\nabla _a,\nabla _b] v^d =R_{abc}{}^dv^c$.  
The curvature on spinors is $[\nabla _a,\nabla _b]\chi ^D=R_{abC}{}^D\chi ^C$.  
Factors of the speed of light, and of Planck's constant $\hbar$, are omitted.

\section{Spinor geometry}

This section reviews spinor calculus and geometry.  The aim is not to give a general exposition,  but to recall the formulas and interpretations which will be used later.  

\subsection{Spinor algebra and geometry}

We assume on physical grounds that a space--time $(M,g_{ab})$ is oriented (with volume-form $\epsilon _{abcd}$) and time-oriented and comes equipped with a spin-structure, by which we mean a complex two-dimensional vector bundle $\Spin ^A$ of (unprimed, right-handed) spinors (the superscript $A$ indicates the index structure) equipped with a skew form $\epsilon _{AB}$ and Hermitian Infeld--van der Waerden symbol $\sigma _a{}^{AA'}$ such that 
\begin{eqnarray}
 g_{ab}&=&\sigma _a{}^{AA'}\sigma _b{}^{BB'}\epsilon _{AB}\epsilon _{A'B'}\\
 \epsilon _{abcd}&=&\sigma _a{}^{AA'}\sigma _b{}^{BB'}\sigma _c{}^{CC'}\sigma _d{}^{DD'}
 (i\epsilon_{AC}\epsilon_{BD}\epsilon _{A'D'}\epsilon _{B'C'}
   -i\epsilon _{AD}\epsilon _{BC}\epsilon _{A'C'}\epsilon _{B'D'}) .\quad
\end{eqnarray}
Downstairs indices correspond to dual quantities and primed indices to complex conjugates, so ${\overline\alpha}^{A'}\in\Spin ^{A'}_p$ is the conjugate of $\alpha ^A\in\Spin ^A_p$.  (By convention, for certain standard fields bars are omitted on complex conjugates; thus $\epsilon _{A'B'}$ is the conjugate of $\epsilon _{AB}$.)
We let $\epsilon ^{AB}$ be minus the inverse of $\epsilon _{AB}$, and use these two spinors to raise and lower indices with the conventions $\alpha ^A=\epsilon ^{AB}\alpha _B$, $\alpha _A =\alpha ^B\epsilon _{BA}$.  
The covariant derivative extends to spinor fields with $\nabla _a\epsilon _{BC}=0$, $\nabla _a\sigma _b{}^{BB'} =0$.  Then any non-zero spinor $\alpha ^A$ determines a null vector $\sigma ^a{}_{AA'}\alpha ^A{\overline\alpha}^{A'}$.  Such vectors will, by continuity, either be all future-pointing or all past-pointing; we assume that they are future-pointing.

Any vector $v^a$ will have a spinor equivalent $v^{AA'} =v^a\sigma _a{}^{AA'}$.  
A skew bivector, that is, a tensor $F^{ab}=-F^{ba}$, will have a spinor equivalent of the form
\begin{equation}
  F^{AA'BB'}=F^{ab}\sigma _a{}^{AA'}\sigma _b{}^{BB'} =\phi ^{AB}\epsilon ^{A'B'}
  +{\psi}^{A'B'}\epsilon ^{AB}\, ,
\end{equation}
where $\phi ^{AB}$, ${\psi}^{A'B'}$ are symmetric, and ${\psi}^{A'B'}  ={\overline\phi}^{A'B'}$ if $F^{ab}$ is real, since any spinor skew on two indices must be proportional to the corresponding $\epsilon ^{AB}$.  We have
\begin{equation}
{}^*F^{AA'BB'} = (1/2)\epsilon ^{AA'BB'}{}_{CC'DD'}F^{CC'DD'}
=-i\phi ^{AB}\epsilon ^{A'B'}
  +i{\psi}^{A'B'}\epsilon ^{AB}\, .
\end{equation}
The parts $F^{AA'BB'}_{-}=\phi ^{AB}\epsilon ^{A'B'}$, $F^{AA'BB'}_{+}=\psi ^{A'B'}\epsilon ^{AB}$ are called the anti-self-dual and self-dual portions of $F^{ab}$.

Any spinor $\alpha ^A$, as already noted, will determine a null vector $\sigma ^a{}_{AA'}\alpha ^A{\overline\alpha}^{A'}$, but that null vector determines the spinor only up to phase.  We may however recover the spinor up to sign from the anti-self-dual skew bivector $F^{ab}=\sigma ^a{}_{AA'}\sigma ^b{}_{BB'}\alpha ^A\alpha ^B\epsilon ^{A'B'}$.  
This is a complex two-plane element (that is, can be written in the form $v^aw^b-w^av^b$), and this two-plane is null (has $\sigma ^a{}_{AA'}\alpha ^A{\overline\alpha}^{A'}$ as a tangent).  Conversely, 
an anti-self-dual bivector $\phi ^{AB}\epsilon ^{A'B'}$ is a two-plane element iff $\phi ^{AB}=\alpha ^A\alpha ^B$ for some $\alpha ^A$ (unique up to sign).
Thus the spinors $\Spin ^A_p$ at any point $p$ form a two-to-one covering of the space of anti-self-dual two-planes.  In this way, spinors are very nearly determined from tensors.

Let us now understand the geometry in terms of a particular frame.  Suppose a unit future-directed timelike vector $t^a$ has been chosen.  Let us write $l^a=\sigma ^a{}_{AA'}\alpha ^A{\overline\alpha}^{A'}$.  We will then have $l^a =E(t^a+z^a)$ for some $E> 0$ (if $\alpha ^A\not=0$) and some unit space-like vector $z^a$ orthogonal to $t^a$.  A bivector $F^{ab}$ containing this will have the form $E(l^av^b-v^al^b)$ for some $v^b$ (unique up to a multiple of $l^a$).  Now the condition $F^{[ab}F^{c]d}=0$ that the bivector have this form will imply, if the bivector is self- or anti-self-dual, that $F^{ab}F_{bc}=0$, and this will mean that $v^al_a=0$.  We may thus use the freedom in $v^a$ to take it to be an element of the two-plane orthogonal to $t^a$ and $z^a$, say $v^a=Ax^a+By^a$ for $(t^a, x^a, y^a, z^a)$ forming an oriented, time-oriented orthonormal basis, and $x^a$, $y^a$ determined uniquely up to a rotation about $z^a$.  
To fix the overall scale of $v^a$, note that 
$(F^{ab}t_b)({\overline F}_{ac}t^c) =-\alpha ^A\alpha ^B t_B{}^{A'}
{\overline\alpha}_{A'}{\overline\alpha}_{C'} t^C{}_{A'}
=-(t_{AA'}\alpha ^A{\overline\alpha}^{A'})^2=-E^2$; but also $F^{ab}t_b =-Ev^a$, so 
\begin{equation}\label{normeq}
v^a{\overline v}_a =-1\, .
\end{equation}

Then we will have
\begin{equation}
 {}^*F^{ab}=2E\left({}^*l^{[a} (Ax^{b]}+By^{b]})\right)  =2El^{[a}(Ay^{b]} -Bx^{b]})\, ,
\end{equation} 
and this will be anti-self-dual iff $A=- iB$ 
and self-dual iff $A=iB$.
Thus $v^a =A(x^a\pm iy^a)$, and by eq. (\ref{normeq}) we will have $|A|=2^{-1/2}$.  There will thus be a unique allowable choice of $x^a$, $y^a$ with $A=2^{-1/2}$.  The two-plane element representing the spinor will then have the form
\begin{equation}\label{spinasd}
  \sqrt{2} E (t^{[a}+z^{[a})(x^{b]} +iy^{b]})\, ,
\end{equation}
for unique $x^a$, $y^a$, $z^a$ once $t^a$ is chosen.
(In (\ref{spinasd}), the square brackets within each set of round parentheses are not nested, as they apply to different terms:  the quantity on this line is the skew part of $\sqrt{2} E (t^{a}+z^{a})(x^{b} +iy^{b})$.)
Note that the real part of this would determine the imaginary part.  Also a multiplication of the spinor $\alpha ^A$ by a phase $e^{i\phi}$ will multiply $v^a$ by $e^{2i\phi}$, and hence rotate $x^a$ and $y^a$ by $+2\phi$ about $z^a$.  (For {\em self}-dual two-planes, the rotation would be by $-2\phi$.)  In this sense the anti-self-dual two-plane, and the spinor $\alpha ^A$, are right-handed.  Conjugate spinors give rise to self-dual two-planes and are left-handed.

The null vector $l^a$ is called the {\em flagpole} of the spinor, and the real half two-plane $\{ Pl^a +Qx^a\mid P\in\R,$ $Q>0\}$ its {\em flag plane}.  Together these are the {\em null flag} which (with knowledge of which chirality is wanted) determine the spinor up to sign.
Alternatively, we could regard the spinor as determined by the real part of eq. (\ref{spinasd}), or the collection of pairs $(Pl^a, Qx^a)$ with $PQ=E$ (or $\sqrt{2}E$) and $P,Q\geq 0$ (and the chirality).  These latter choices are analytically a bit more natural, for they more explicitly integrate the complex structure.

We thus identify the spinors at a point with the two-to-one covering of the set of anti-self-dual two-planes at the point.

\subsection{Conformal rescalings}

We have just reviewed the geometric interpretation of spinors, and seen that, up to a sign, a spinor can be regarded in either of two ways:  as an anti-self-dual two-plane element (or the real part of that); or as a null flag.  Each of these concepts is clearly invariant under conformal rescalings, and so either can be used as a basis for determining the behavior of spinors in response to them.  

There is, however, a bit of a subtlety:  while each of the methods described of representing a spinor geometrically is conformally invariant, the translation between the two is not.  This can be seen from eq. (\ref{spinasd}), where to convert the null flag to a two-plane element, one must wedge the flagpole $E(t^a+z^a)$ with an appropriate complex vector {\em normalized with respect to the metric}.  
This means that the two definitions, while equivalent with any one choice of scale for the metric, will have different behaviors under conformal rescalings.

(There is in principle a further point.  What we have shown is that, with either definition, the {\em set} of spinors is conformally invariant, but we have not shown its vector-space structure is.  It is in fact clear geometrically that, with either definition, multiplication by complex scalars is conformally invariant.  Penrose and Rindler \cite{PR1984} give an explicit geometric treatment of addition of spinors viewed as flag planes, which clearly does not depend on the scale of the metric.  For spinors viewed as two-plane elements, we may give a simple computational argument:
Suppose
$F^{AA'BB'}=\alpha ^A\alpha ^B\epsilon ^{A'B'}$, $G^{AA'BB'}=\beta ^A\beta ^B\epsilon ^{A'B'}$ represent spinors.  Then a direct calculation shows that
\begin{equation}
  F^{ab} \pm 4(F^{c[a}G^{b]d} g_{cd} )( F^{pq}G^{rs}g_{pr}g_{qs}) ^{-1/2} +G^{ab}
\end{equation}  
represents the spinor $\alpha ^A +\beta ^A$ (or $\alpha ^A-\beta ^A$).  This is evidently conformally invariant.)

The choice in definition of spinor, and the concomitant conformal rescaling properties, are for many purposes (including here) essentially matters of convenience.  However, in certain circumstances, the difference could be of physical significance.  For instance, in a path-integral attempt to quantize gravity coupled to Fermions, in general one would expect different results from the different choices.

{\em If we identify spinors with the two-to-one covers of null flags,} then because any spinor $\alpha ^A$ determines a flagpole $\sigma ^a{}_{AA'}\alpha ^A{\overline\alpha}^{A'}$, the Infeld--van der Waerden symbol $\sigma ^a{}_{AA'}$ (and its inverse $\sigma _a{}^{AA'}$) will have conformal weight zero, and the interconversion of tensor and spinor indices will commute with conformal rescaling.  Because the argument of $\epsilon _{AB}\alpha ^A\beta ^B$ is conformally invariant \cite{PR1984}, the spinor $\epsilon _{AB}$ can change only by a real factor, and then the 
equation $g_{AA'BB'}=\epsilon _{AB}\epsilon _{A'B'}$ forces $\epsilon _{AB}$ to have conformal weight $+1$.  These are the choices made in Penrose and Rindler \cite{PR1984,PR1986}.

{\em If we identify spinors with the two-to-one covers of anti-self-dual two-plane elements,} the identity
\begin{equation}\label{epseq}
  \epsilon _{AB}\alpha ^A\beta ^B =\pm (1/2)(F^{ab}G^{cd} g_{ac}g_{bd})^{1/2}\, ,
\end{equation}
requires us to assign $\epsilon _{AB}$ conformal weight $+2$.  Then $\epsilon ^{AB}$ will have conformal weight $-2$, and from the equation
\begin{equation}\label{sigeq}
  F^{ab}\sigma _a{}^{AA'}\sigma _b{}^{BB'} =\alpha ^A\alpha ^B\epsilon ^{A'B'}
\end{equation}
we see we must assign $\sigma _a{}^{AA'}$ weight $-1$.

{\em More generally, we may consider a weighted geometric average of these two.}  Taking a null flag modulo its flagpole  gives a half-line (in the example in the previous subsection, this would be the $+x^a$ direction); the metric descends to this space.  If we take a spinor to be an anti-self-dual two-plane element tensored with an element of the $w^{\rm th}$ power of that space, then the case $w=0$ corresponds to the anti-self-dual two-plane elements and the case $w=-1$ to the null flags.  
(In principle we could take $w$ to be any complex number, but we will assume $w$ is real.)
Explicitly, and element of the tensor power has the form $F^{ab}\otimes (X^c)^w$, where $X^c$ would be a vector in the direction $+x^c$ in the notation of the previous subsection.  We reduce this to a two-plane element, with respect to a particular metric, by writing $X^c=\xi x^c$ where $\xi ^2=-g_{ab}X^cX^d$, and mapping $F^{ab}\otimes (X^c)^w \mapsto \xi ^w F^{ab}$.  We see that $\xi$ has conformal weight $+1$, so $\xi ^w F^{ab}$ has conformal weight $+w$.  Then equation (\ref{epseq}) (with a factor of $\xi ^w$ appearing with $F^{ab}$, and a corresponding factor for $G^{ab}$) shows $\epsilon _{AB}$ has conformal weight $+2+w$, and equation (\ref{sigeq}) shows $\sigma _a{}^{AA'}$ has weight $-1-w$.

The choice $w=-2$ corresponds to the work of Kosmann \cite{Kosmann1971} and those following it (and was also used in earlier work of Penrose \cite{RP1965}, for different reasons).  It has the advantage of giving $\epsilon _{AB}$ conformal weight zero (and this corresponds in Kossman's approach to taking the Dirac $\gamma _a$'s as fundamental);
it can also be viewed as giving $F_a{}^b$ conformal weight zero, or as identifying spinors with elements of the covering space of those infinitesimal Lorentz transformations corresponding to null rotations.

\subsection{Four-spinors}

A Dirac spinor is a pair
\begin{equation}
\psi =\left[\begin{matrix} \psi ^{Q'}\\ \psi _Q\end{matrix}\right]
\end{equation}
of a conjugate and a dual spinor.  The Dirac adjoint is
\begin{equation}
\tilde\psi =\left[\begin{matrix} {\overline\psi}_{Q'}
  &{\overline\psi}^Q\end{matrix}\right]\, ,
\end{equation}
and the Dirac gamma is given by
\begin{equation}
\gamma _a=\sqrt{2}\left[\begin{matrix}
  0&\sigma _a{}^{P'Q}\\ \sigma _{aPQ'}&0\end{matrix}\right]
\end{equation}
and satisfies $\gamma _a\gamma _b+\gamma _b\gamma _a=2g_{ab}$.  
The conventions are chosen here to be compatible with those usual in relativity and also quantum field theory.  With the choices here,
the standard basis described on pp. 120--125 of Penrose and Rindler \cite{PR1984} or pp. 6--8 of \cite{PR1986} gives precisely the Weyl basis on p. 79 of Schweber \cite{Schweber1961}.
(As noted above, these conventions for the {\em relation} between two- and four-component spinors, for the Clifford algebra, and for the Dirac gamma, differ from those  of Penrose and Rindler, but that will not matter for us.)

While the formulas just given are the only ones we will need here, we note for convenience of comparison with the quantum fields literature a few further ones.
We take
\begin{equation}
\gamma _5=\frac{1}{24}\epsilon _{abcd}\gamma ^a\gamma ^b\gamma ^c\gamma ^d =\left[\begin{matrix} i&\\&-i\end{matrix}\right]\, ,
\end{equation}     
where blank places are occupied by zeroes.
(Many authors use $-i$ times this.)  The projectors to the left- (respectively, right-) handed spinors are 
\begin{equation}
(1/2)\left[ 1- i\gamma _5\right]
=\left[\begin{matrix} 1&\\&0\end{matrix}\right]\, 
   \quad\text{and}\quad
(1/2)\left[ 1+ i\gamma _5\right]
  =
 \left[\begin{matrix} 0&\\&1\end{matrix}\right]\, .
\end{equation}

Finally, the charge conjugate of a spinor is
\begin{eqnarray}
  \psi _{\rm c}&=&\left[\begin{matrix}0&\epsilon ^{Q'R'}\\
    \epsilon _{QR}&0\end{matrix}\right] \left[\begin{matrix}
      {\overline\psi} ^R\\ {\overline\psi}_{R'}\end{matrix}
      \right]\nonumber\\
  &=&\left[\begin{matrix} {\overline\psi}^{Q'}\\
    {\overline\psi}_Q\end{matrix}\right]\, .
\end{eqnarray}    
(Many quantum field theory books give an apparently basis-dependent expression for this, in terms of $\gamma _2$; the formula here makes the invariance clear.
It is worth remarking that this formula, and the discussion of the previous subsection, also make clear that charge conjugation does {\em not} commute with conformal rescaling unless $w=-2$.)

\section{Lie derivatives of spinor fields}

In this section, we see how a natural definition of the Lie derivative of a spinor field may be given.  The ideas are very close to those of Kosmann, but the connection with the space--time geometry of spinors is emphasized.
The aims are to get a useful calculus and an explicit geometric justification of it, and so it will be helpful to begin with some basic points.

In general, differentiating a vector-bundle-valued quantity along a curve is problematic because, in trying to form the numerator of the difference quotient, one needs to compare elements of two different spaces, the fibres over the two points in question.  If the curve is one of the integral curves of a vector field $\xi ^a$, one can use the associated flow $\Phi _t$ to push and pull elements of the tensor algebra along the field, and the differentiation one gets from this is the Lie derivative. 

Let $\xi ^a$ be a vector field on space--time generating a local flow $\Phi _t$, so $\partial _t\Phi (p)=\xi ^a(p)$.\footnote{In the abstract-index notation, the flow does not carry a tensor index.  In a coordinate representation, it would.}  Then we get a one-parameter family of motions on the tensor algebra.  For any scalar field $\phi$, we have
\begin{equation}
  \phi _t(p)=\phi (\Phi _t(p))\, ,
\end{equation}
that is, a one-parameter family of values at any fixed $p$.  Then
$\partial _t\phi _t =\xi^a\nabla _a\phi$, the Lie derivative of $\phi$ along $\xi ^a$.

For vector fields, we make use of the push-forward
\begin{equation}
  (\Phi _{t*})^a{}_b:T_p(M)\to T_{\Phi _t(p)}(M)
\end{equation}
from the tangent space at $p$ to that at $\Phi _t(p)$.  Then for
any vector field $v^a$, we define
\begin{equation}
  v^a_t(p) =(\Phi _{t*}^{-1})^a{}_bv^b(\Phi _t(p))\, .
\end{equation}
For fixed $p$, this is a one-parameter family of vectors, and
$\partial _tv^a _t =\xi ^b\nabla _b v^a -v^b\nabla _b\xi ^a$ is the Lie derivative of $v^a$ along $\xi^a$.  The action on the remainder of the tensor algebra is determined by linearity and the Leibniz rule, and the derivatives with respect to $t$ give the corresponding Lie derivatives.

In what follows, we will have tensors transported from $\Phi _t(p)$ to $p$ via appropriate tensor products of the inverse push-forward and the pull-back.  Because of the many indices involved, it will be convenient not to use the subscript $t$ (as we did for $v_t^a(p)$) for such quantities, but simply to indicate them by hats (thus ${\hat v}^a=v^a_t(p)$).

To construct a corresponding action on spinors, we wish to identify the spin space $\Spin ^A_{\Phi _t(p)}(M)$ at $\Phi _t(p)$ with that $\Spin ^A_p(M)$ at $p$.  The inverse push-forward gives an isomorphism of the corresponding tangent spaces, and hence takes two-plane elements at $\Phi _t(p)$ to those at $p$.  However, it will take a two-plane element $F^{ab}(\Phi _t(p))$ anti-self-dual with respect to $g_{ab}(\Phi _t(p))$ to ${\hat F}^{ab}$ which is anti-self-dual with respect to ${\hat g}_{ab}$ rather than $g_{ab}(p)$, and so it will not directly determine an element of $\Spin ^A_p(M)$.
We can remedy this, however, by introducing an appropriate, if Lie-artificial, linear transformation.  

For each $t$, the quantity 
$g^{bc}{\hat g}_{ca}$
is a map from 
$T_p(M)$
to itself, varying smoothly with $t$ and equal to the identity at $t=0$.  It thus will have, for $t$ close enough to zero, a unique square root close to the identity, say ${\hat M}^b{}_{a }$.  In fact, we can use the expansion $\sqrt{ 1+x} =1+x/2+\cdots$ to give ${\hat M}^b{}_{a }$ as a sum of powers of $ {\hat h}^b{}_a=
g^{bc}{\hat g}_{ca} -\delta ^b_a$ times numerical coefficients
(for $t$ close to zero).  
Now ${\hat h}^b{}_a$ is clearly ${\hat g}^{ab}$-symmetric, that is, the form 
${\hat h}^a{}_c {\hat g}^{cb} =g^{ab}$
is symmetric.
Similarly any power of ${\hat h}^b{}_a$ will be ${\hat g}^{ab}$-symmetric.
Thus ${\hat M}^b{}_{a }$ will also be ${\hat g}^{ab}$-symmetric:
\begin{equation}
  {\hat M}^b{}_{c}{\hat g}^{ca} ={\hat g}^{ba}{\hat M}^c{}_{a }\, .
\end{equation}  

Now we will have
\begin{eqnarray}
  \epsilon ^{ab}{}_{pq} {\hat M}^p{}_c {\hat M}^q{}_d 
  &=&g^{am} g^{bn} \epsilon _{mnpq}
    {\hat M}^p{}_c {\hat M}^q{}_d  \nonumber\\
  &=& {\hat M}^a{}_r {\hat M}^r{}_k{\hat g}^{km}
          {\hat M}^b{}_s{\hat M}^s{}_l {\hat g}^{ln}
    \epsilon _{mnpq}
    {\hat M}^p{}_c {\hat M}^q{}_d  \nonumber\\
  &=&{\hat M}^a{}_r {\hat M}^m{}_k{\hat g}^{kr}
          {\hat M}^b{}_s{\hat M}^n{}_l {\hat g}^{ls}
    \epsilon _{mnpq}
    {\hat M}^p{}_c {\hat M}^q{}_d  \nonumber\\
&=&(\det\hat M ){\hat M}^a{}_r{\hat g}^{kr}
 {\hat M}^b{}_s{\hat g}^{ls}
 \epsilon _{klcd}\nonumber\\
 &=&{\hat M}^a{}_r{\hat g}^{kr}
 {\hat M}^b{}_s{\hat g}^{ls}
 {\hat\epsilon} _{klcd}\nonumber\\
  &=&{\hat M}^a{}_r
 {\hat M}^b{}_s
 {\hat\epsilon} ^{rs}{}_{cd}\, .
\end{eqnarray}
This means that ${\hat M}^p{}_c$ (or more properly, the tensor product of that with itself) intertwines the duality operators $(1/2){\hat\epsilon}^{ab}{}_{cd}$ for ${\hat g}_{ab}$ and $(1/2)\epsilon ^{ab}{}_{cd}$ for $g_{ab}$; it will take ${\hat g}_{ab}$-anti-self-dual two-planes to $g_{ab}$-anti-self-dual ones.  We could use this as a basis for identifying spinors.  However, to agree with the behavior we have deduced for the effects of conformal rescalings, we include a power of $\det {\hat M}$.  Precisely, we have a map from the ${\hat g}_{ab}$-anti-self-dual two planes to those with respect to $g_{ab}$ given by
\begin{equation}\label{isom}
  F^{ab}\mapsto  (\det \hat M )^{(-1/2)-(w/4)}{\hat M}^a{}_p {\hat M}^b{}_q F^{pq}\, .
\end{equation}

We now apply this.  Let $F^{ab}$ be a field of anti-self-dual bivectors (they need not be two-plane elements), with the spinor form
\begin{equation}
  F^{AA'BB'}=\phi ^{AB}\epsilon ^{A'B'}\, .
\end{equation}
Let ${\hat F}^{ab}$ be the associated one-parameter family induced by the flow, so that
\begin{equation}
\partial _t{\hat F}^{ab} =\xi ^c\nabla _cF^{ab} -(\nabla _p\xi ^a)F^{pb}
  -(\nabla _p\xi ^b)F^{ap}\, .
\end{equation}
We will define ${\hat\phi}^{AB}$ by
\begin{equation}\label{hatphidef}
{\hat F}^{ab}\sigma _a{}^{AA'}\sigma _b{}^{BB'}
  ={\hat\phi}^{AB}\epsilon ^{A'B'}\, .
\end{equation}
Note here that neither the Infeld--van der Waerden symbol nor the alternating spinor carry hats, as the isomorphism (\ref{isom}) which has been defined represents the result of carrying a $w$-weighted ${\hat g}_{ab}$-anti-self-dual bivector at $\Phi _t(p)$  to a $g_{ab}$-anti-self-dual one at $p$, and this weighted bivector is {\em identified} with the spinor $\phi ^{AB}$.  The Infeld--van der Waerden symbol, and the skew form, in equation (\ref{hatphidef}) simply serve as conveniences to relate ordinary tensor quantities at $p$ to a spinor ${\hat\phi}^{AB}$.

We wish to compute
\begin{equation}
\partial _t \left( (\det \hat M )^{(-1/2)-(w/4)} {\hat M}^a{}_p{\hat M}^b{}_q{\hat F}^{pq}\right)
\end{equation}
(at $t=0$).  We have $\partial _t{\hat M}^a{}_p =(1/2)(\nabla ^a\xi _p+\nabla _p\xi ^a)$ and hence $\partial _t(\det \hat{M}) =\nabla\cdot\xi$.  Then
\begin{eqnarray}
\partial _t \left( (\det \hat M)^{(-1/2)-(w/4)} {\hat M}^a{}_p{\hat M}^b{}_q{\hat F}^{pq} \right)
&=&\xi\cdot\nabla F^{ab} -(\nabla _p\xi ^a)F^{pb}
  -(\nabla _p\xi ^b)F^{ap} \nonumber\\
  &&+(1/2)(\nabla ^a\xi _p +\nabla _p\xi ^a) F^{pb}
  +(1/2)(\nabla ^b\xi _p +\nabla _p\xi ^b) F^{ap}\nonumber\\
&&  +((-1/2)-(w/4))(\nabla\cdot\xi )F^{ab} 
\nonumber\\
&=&\xi\cdot\nabla F^{ab}\\
  &&+(1/2)(\nabla ^a\xi _p -\nabla _p\xi ^a) F^{pb}
  +(1/2)(\nabla ^b\xi _p -\nabla _p\xi ^b) F^{ap}
  \nonumber\\
&&  +((-1/2)-(w/4))(\nabla\cdot\xi )F^{ab}
\, .
\end{eqnarray}  
The spinor form of this is, after a short calculation,
\begin{equation}\label{twospinder}
\partial _t{\hat\phi}^{AB}=\xi\cdot\nabla \phi ^{AB} +\gamma _P{}^A\phi ^{PB} +\gamma _P{}^B\phi ^{AP}
  +((-1/2)-(w/4))(\nabla\cdot\xi)\phi ^{AB} \, ,
\end{equation}
where
\begin{equation}
\gamma ^{AP} =-(1/2)\nabla ^{(A}{}_{Q'}\xi ^{P)Q'}\, .
\end{equation}  

Equation (\ref{twospinder}) represents the linearization of the association $\phi^{AB}\mapsto {\hat\phi}^{AB}$ defined in eq. (\ref{hatphidef}).  From eq. (\ref{twospinder}) it is evident that this association factors to one-index spinor fields $\chi ^A\mapsto {\hat\chi}^A$ with linearization
\begin{equation}\label{spLiedef}
\partial _t{\hat\chi}^A =\xi \cdot\nabla \chi ^A +\gamma _P{}^A\chi ^P
  +((-1/4)-(w/8))(\nabla\cdot\xi)\chi ^A\, .
\end{equation}  
At this point, the operator $\partial _t$ (composed with hatting), representing Lie differentiation, can be defined on arbitrary tensor products of tensors and spinors, by linearity and the Leibniz rule.  Besides the initial covariant derivative operator $\xi\cdot\nabla$, there will be one $\nabla _a\xi ^b$ term for each tensor index, according to the usual rule (negative contributions for upstairs indices and positive for downstairs), and similarly contributions like eq. (\ref{spLiedef}) for spinor indices (with negative contributions for dual spinors).  We now replace the notation $\partial _t$ (combined with the hat) with $\Lie _\xi$.

Some discussion of these formulas and their consequences is in order.  First, note that $\gamma ^{AP}$ is essentially the anti-self-dual part of $\nabla ^{[a}\xi ^{p]}$, and hence is sensitive, in metric terms, to infinitesimal Lorentz motions (usually called the rotation) induced by flow along $\xi ^a$.  The other term, proportional to the divergence $\nabla\cdot\xi$, is sensitive to infinitesimal changes of conformal scale, but neither of these terms shows a dependence on the shear, the trace-free part of $\nabla _{(a}\xi _{b)}$, which would represent an infinitesimal distortion of the conformal structure.  This is very different from the formulas for Lie derivatives of tensor quantities, which would involve such dependences.

The reason for this is that the flow along a vector field distorts those elements of the tensorial geometry which represent spinors to ones which do not --- for instance, converting anti-self-dual two-plane elements to two-plane elements which are not eigenquantities of the duality operator (or, from the point of view of Penrose and Rindler, taking null flags to non-null half-planes).
In other words, the natural Lie derivative of a spinor would not be a spinor at all, but some more general quantity (taking values in a covering space of the space of the two-plane elements).  We have, however, by somewhat artificial means shoe-horned it into the space of spinors, and the resulting `Lie derivative' has terms incompatible with this projected out.

One might think at this point that so much geometry has been discarded in passing to the spinor Lie derivative that its significance is very limited.  
We can recover the missing geometry easily enough, however, by recalling that the relation (\ref{sigeq}) between spinors and tensors is set by the Infeld--van der Waerden symbol $\sigma _a{}^{AA'}$ and the alternating spinor $\epsilon _{AB}$.  It is the Lie derivatives of these quantities, then, which code this information.  Because they have been constructed to satisfy the Leibniz rule and are therefore compatible with ordinary Lie differentiation of tensors, and because spinors are recoverable from tensors up to a sign, no differential-geometric information is lost if we keep track of $\Lie _\xi \sigma _a{}^{AA'}$ and $\Lie _\xi \epsilon _{AB}$.

We have
\begin{equation}
\Lie _\xi\epsilon _{AB}= ((1/2)+(w/4))(\nabla\cdot\xi )\epsilon _{AB}\, ,
\end{equation} 
and this is compatible with the conformal scalings discussed earlier, with $(1/4)\nabla\cdot\xi$ representing the infinitesimal change in conformal factor.
(So with Kosmann's definition, corresponding to $w=-2$, we have $\Lie _\xi \epsilon _{AB}=0$; Penrose--Rindler spinors, with $w=-1$, would have $\Lie _\xi \epsilon _{AB}=(1/4)(\nabla\cdot\xi) \epsilon _{AB}$.)

Now let us consider
\begin{eqnarray}
\Lie _\xi\sigma _a{}^{BB'} &=&(\nabla _a\xi ^p)\sigma _p{}^{BB'}
  +\gamma _P{}^B\sigma _a{}^{PB'} +\gamma _{P'}{}^{B'}\sigma _a{}^{BP'}
   +((-1/2)-(w/4))(\nabla\cdot\xi ) \sigma _a{}^{BB'}\nonumber\\
   &=&\nabla _a\xi ^{BB'} 
  +\gamma _P{}^B\sigma _a{}^{PB'} +\gamma _{P'}{}^{B'}\sigma _a{}^{BP'}
   +((-1/2)-(w/4))(\nabla\cdot\xi ) \sigma _a{}^{BB'}\, .
\end{eqnarray} 
This is most easily computed by passing to its spinor form and lowering indices:
\begin{eqnarray}
&&\nabla _{AA'}\xi _{BB'} 
  +\gamma _{AB}\epsilon_{A'B'} +\gamma _{A'B'}\epsilon _{AB}
   +((-1/2)-(w/4))(\nabla\cdot\xi ) \epsilon _{AB}\epsilon _{A'B'}\nonumber\\
   &=&\tau _{AA'BB'} +((1/4)-(w/4))(\nabla\cdot\xi )\epsilon _{AB}\epsilon _{A'B'}\, ,
\end{eqnarray} 
where 
\begin{equation}
  \tau _{ab}=\nabla _{(a}\xi _{b)} -(1/4)g_{ab}\nabla\cdot\xi
\end{equation}
is the trace-free and symmetric part of $\nabla _a\xi _b$.  Thus
\begin{equation}
\Lie _\xi\sigma _a{}^{BB'} =\tau _a{}^{BB'} +((-1/4)-(w/4))\sigma _a{}^{BB'}\nabla\cdot\xi\, .
\end{equation}     
We see here explicitly that this captures the shear (and divergence, if $w\not= -1$) of the vector field $\xi ^a$.

We close this section with some further results which will be useful.
We have
\begin{eqnarray}
[\Lie _\xi,\nabla _b]\chi ^A
&=&\xi\cdot\nabla\nabla _b\chi ^A
 +(\nabla _b\xi ^c)\nabla _c\chi ^A
 +\gamma _P{}^A\nabla _b\chi ^P
 -((1/4)+(w/8))(\nabla\cdot\xi )\nabla _b\chi ^A\nonumber\\
&& -\nabla _b\left(\xi\cdot\nabla \chi ^A
   +\gamma _P{}^A\chi ^P -((1/4)+(w/8))(\nabla\cdot\xi)\chi ^A\right)\nonumber\\
 &=& \xi ^cR_{cbP}{}^A \chi ^P 
   -(\nabla _b\gamma _P{}^A)\chi ^P
   +((1/4)+(w/8))(\nabla _b\nabla\cdot\xi)\chi ^A\, ,
\end{eqnarray}   
and likewise
\begin{equation}\label{Liecomm}
  [\Lie _\xi,\nabla _b]\chi _A 
=-\xi ^cR_{cbA}{}^P \chi _P 
   +(\nabla _b\gamma _A{}^P)\chi _P
   -((1/4)+(w/8))(\nabla _b\nabla\cdot\xi)\chi _A\, .
\end{equation}   
Therefore, we have
\begin{eqnarray}
[\Lie _\xi,\sigma ^b{}_{BB'}\nabla _b]\chi ^A
 &=&(\Lie _\xi\sigma ^b{}_{BB'})\nabla _b\chi ^A
   +\sigma ^b{}_{BB'}[\Lie _\xi,\nabla _b]\chi ^A\nonumber\\
&=&(-\tau ^b{}_{BB'} +((1/4)+(w/4))(\nabla\cdot\xi) \sigma ^b{}_{BB'}
 )\nabla _b\chi ^A
 +
 \xi ^cR_{cBB'P}{}^A \chi ^P \nonumber\\
 &&
   -(\nabla _{BB'}\gamma _P{}^A)\chi ^P
   +((1/4)+(w/8))(\nabla _{BB'}\nabla\cdot\xi)\chi ^A
\end{eqnarray}

\section{Application to Lagrangians}

In this section, we apply the formalism developed previously to find the stress--energies for spinor theories.  
Let
\begin{equation}
\psi =\left[\begin{matrix} \psi ^{A'}\\ \psi _A\end{matrix}
\right]
\end{equation}
be a spinor field whose components we take to be Grassmann quantities.
We consider a Lagrangian
\begin{equation}
\LL =\LL _{\rm kinetic}+\LL _{\rm mass}
\end{equation}
where the kinetic term is
\begin{eqnarray}
 \LL _{\rm kinetic} &=& (i/2){\tilde\psi} \gamma\cdot \nabla\psi
    -(i/2)(\nabla _a{\tilde\psi})\gamma ^a\psi\nonumber\\
    &=& i2^{-1/2} ({\overline\psi}_{A'}\nabla ^{AA'}\psi _A
     +{\overline\psi}^A\nabla _{AA'}\psi ^{A'}
     -(\nabla ^{AA'}{\overline\psi}_{A'})\psi _A
     -(\nabla _{AA'}{\overline\psi}^A)\psi ^{A'})\, ,\qquad
\end{eqnarray}
and the mass term
\begin{eqnarray}
\LL_{\rm mass}    
  &=&     -m({\overline\psi}_{A'}\psi ^{A'} +{\overline\psi}^A \psi _A )\nonumber\\
&&   -\mA (\epsilon ^{AB}\psi _B\psi _A +\epsilon ^{A'B'}
   {\overline\psi}_{A'}{\overline\psi}_{B'})
   -\mB (\epsilon _{B'A'}\psi ^{B'}\psi ^{A'}
    +\epsilon _{BA}{\overline\psi}^A{\overline\psi}^B)
\end{eqnarray}
allows both Dirac and Majorana contributions; here $m$, $\mA$, $\mB$ are independent real numbers.   

The field equations, determined by requiring the variation of the action with respect to $\psi$ to vanish, are:
\begin{eqnarray}
i2^{1/2}\nabla ^{AA'}\psi _A -m\psi ^{A'} -2\mA {\overline\psi}^{A'}
  &=&0\\
i2^{1/2}\nabla _{AA'}\psi ^{A'} -m\psi _A -2\mB {\overline\psi}_A
  &=&0\, .
\end{eqnarray}  
These equations are valid in flat or curved space--time.  Let us note that, when the field equations are satisfied, we have
\begin{eqnarray}
\LL &=&{\overline\psi}_{A'}(+2\mA {\overline\psi}^{A'})
  +{\overline\psi}^A( +2\mB {\overline\psi}_A)
  -\mA (\psi ^A\psi _A +{\overline\psi}_{A'}{\overline\psi}^{A'})
  -\mB (\psi _{A'}\psi ^{A'}+{\overline\psi}^A{\overline\psi}_A)
  \nonumber\\
&&  -i2^{-1/2}\nabla ^{AA'}({\overline\psi}_{A'}\psi _A)
  -i2^{-1/2}\nabla _{AA'}({\overline\psi}^A\psi ^{A'})
  \nonumber\\
  &=&\mA (\psi ^A\psi _A -{\overline\psi}_{A'}{\overline\psi}^{A'})
  +\mB (-\psi _{A'}\psi ^{A'}+{\overline\psi}^A{\overline\psi}_A)\nonumber\\
&&  -i2^{-1/2}\nabla ^{AA'}({\overline\psi}_{A'}\psi _A)
  -i2^{-1/2}\nabla _{AA'}({\overline\psi}^A\psi ^{A'})
\, .
\end{eqnarray}
This must vanish, by Hermiticity.  Thus 
\begin{equation}
 \LL =0
\end{equation}
and we have the identity
\begin{equation}
i2^{-1/2}\nabla ^{AA'}\left({\overline\psi}_{A'}\psi _A +{\overline\psi}_A\psi _{A'} \right)
=\mA (\psi ^A\psi _A -{\overline\psi}_{A'}{\overline\psi}^{A'})
  +\mB (-\psi _{A'}\psi ^{A'}+{\overline\psi}^A{\overline\psi}_A)
\, .
\end{equation}
Because the Lagrangian vanishes when the field equations are satisfied, the contribution to the stress--energy from the variation of the volume form will vanish.

We now consider the response of the Lagrangian to an infinitesimal diffeomorphism generated by a vector field $\xi ^a$.  This will be specified by applying the Lie derivative $\Lie _\xi$, as usual, but now we may use the spinor Lie derivative to break down each term.  We will
discard any terms which will vanish `on-shell' (that is, when the field equations are satisfied), and the terms coming from simply replacing $\psi ^{A'}$, $\psi _A$ by their Lie derivatives will fall in this class.  
For this reason also, the choice of whether we regard $\psi ^{A'}$ as a spinor or as an index-raised dual spinor $\psi ^{A'}=\epsilon ^{A'B'}\psi _{B'}$, or by fiat give it some conformal weight, will have no affect on the stress--energy.  (And similarly for $\psi _A$.)  
Therefore we shall take what seem the most natural conventions, that $\psi ^{A'}$ is a conjugate spinor and $\psi _A$ is a dual spinor.
If we were in the usual, Bosonic or classical case, the remaining terms would come simply from variations of the metric.  Here we have
\begin{eqnarray}
\Lie _\xi\LL 
&=&i2^{-1/2}\left\{ {\overline\psi}_{A'} [\Lie _\xi,\nabla ^{AA'}]\psi _A
  +{\overline\psi}^A[\Lie _\xi ,\nabla _{AA'}]\psi ^{A'} \right.\nonumber\\
&&\qquad\left.  -[\Lie _\xi\nabla ^{AA'},{\overline\psi}_{A'}] \psi _A
  -[\Lie _\xi ,\nabla _{AA'}{\overline\psi}^A]\psi ^{A'} \right\}\nonumber\\
  &&+((1/2)+(w/4))(\nabla\cdot\xi ) \left(
    \mA (\epsilon ^{AB}\psi _B\psi _A +\epsilon ^{A'B'}{\overline\psi}_{A'}{\overline\psi}_{B'})
    \right.\nonumber\\
    &&\qquad\left. -\mB (\epsilon _{B'A'}\psi ^{B'}\psi ^{A'} +\epsilon _{BA}{\overline\psi}^A{\overline\psi}^B)\right)
  +\text{terms vanishing on-shell} .\qquad
\end{eqnarray}  

The commutator terms are initially lengthy but do simplify.  We have
\begin{eqnarray}
i{\overline\psi}_{A'}[\Lie _\xi ,\nabla ^{AA'}]\psi _A
&=&i{\overline\psi}_{A'}[\Lie _\xi ,\sigma ^{bAA'}\nabla _b]\psi _A\nonumber\\
&=&i(\Lie _\xi \sigma ^{bAA'}){\overline\psi}_{A'}\nabla _b\psi _A
  +i\sigma ^{bAA'}{\overline\psi}_{A'}[\Lie _\xi,\nabla _b]\psi _A\, .
\end{eqnarray}
The second of these terms is readily evaluated using equation (\ref{Liecomm}):
\begin{eqnarray}
  i\sigma ^{bAA'}{\overline\psi}_{A'}[\Lie _\xi,\nabla _b]\psi _A
  &=&i{\overline\psi}_{A'}\left(
    -\xi ^{CC'}R_{CC'}{}^{AA'}{}_A{}^P\psi _P
      +(\nabla ^{AA'}\gamma _A{}^P)\psi _P \right.\nonumber\\
&&\left.      +((1/4)+(w/8))(\nabla ^{AA'}\nabla\cdot\xi)\psi _A
      \right)\, .
\end{eqnarray}
Each of these terms is purely anti-Hermitian and will therefore cancel against a corresponding term.  This is clearest for the last term, which is the contraction of the anti-Hermitian $i{\overline\psi}_{A'}\psi _A$ with a real vector.  The first of the three can be written as
\begin{equation} 
  -i{\overline\psi}_{A'}\xi ^{CC'} (\Phi _{C'}{}^{A'}{}_C{}^P  -3 \Lambda \epsilon _{C'}{}^{A'} \epsilon _C{}^P)\psi _P
  =-i{\overline\psi}_{A'}\xi ^{CC'} (\Phi _{C'}{}^{A'}{}_C{}^A  -3 \Lambda \epsilon _{C'}{}^{A'} \epsilon _C{}^A)\psi _A\, ,
\end{equation}  
where $\Phi _{ab}$ is (minus one-half times) the trace-free part of the Ricci tensor and $\Lambda$ is (one twenty-fourth of) the scalar curvature.  This is manifestly anti-Hermitian.
The second term is proportional to
\begin{equation}
2\nabla ^R{}_{A'}\nabla _{(A}{}^{R'}\xi _{R)R'}
= (1/2)\left([\nabla ^R{}_{A'},\nabla _A{}^{R'}]_{+}\right) \xi _{RR'}
  +(1/2)\nabla ^2\xi _{AA'}
  -2\Phi _{ARA'R'}\xi ^{RR'}
    +6\Lambda \xi _{AA'}\, ,
\end{equation} 
where $[\cdot,\cdot ]_{+}$ indicates the anticommutator.
This is Hermitian, and so when contracted with $i{\overline\psi}_{A'}\psi _A$ will give an anti-Hermitian contribution, which must be cancelled by one from another term.  Similar analyses will hold for the other commutator terms.

We therefore have
\begin{eqnarray}
\Lie _\xi\LL &=&i2^{-1/2}\left\{ (\Lie _\xi\sigma ^{bAA'})
\left({\overline\psi}_{A'}\nabla _b\psi
  -(\nabla _b{\overline\psi}_{A'})\psi _A\right)
  +   (\Lie _\xi\sigma ^b{}_{AA'})\left( {\overline\psi}^A\nabla _b\psi ^{A'}
  -(\nabla _b{\overline\psi}^A)\psi _{A'}\right)
  \right\} 
    \nonumber\\
 &&+((1/2)+(w/4))(\nabla\cdot\xi ) \left(
    \mA (\epsilon ^{AB}\psi _B\psi _A +\epsilon ^{A'B'}{\overline\psi}_{A'}{\overline\psi}_{B'})
    \right.\nonumber\\
    &&\qquad\left. -\mB (\epsilon _{B'A'}\psi ^{B'}\psi ^{A'} +\epsilon _{BA}{\overline\psi}^A{\overline\psi}^B)\right)
 +\text{terms vanishing on-shell}\, .      
\end{eqnarray}  
Using
\begin{eqnarray}
  \Lie _\xi\sigma ^{bAA'} &=&
             -\nabla ^{AA'}\xi ^b
             +\gamma _P{}^A\sigma ^{bPA'}
             +\gamma _{P'}{}^{A'}\sigma ^{bAP'}
             -((1/2)+(w/4))(\nabla\cdot\xi)\sigma ^{bAP'}
                 \nonumber\\
               &=&-\tau ^{AA'b}
                            -((3/4)+(w/4))(\nabla\cdot\xi)\sigma ^{bAP'}\, ,\\
  \Lie_\xi\sigma ^b{}_{AA'}&=&
    -\nabla _{AA'}\xi ^b 
    -\gamma _A{}^P\sigma ^b{}_{PA'}
    -\gamma _{A'}{}^{P'}\sigma ^b{}_{AP'}
    +((1/2)+(w/4))(\nabla\cdot\xi )\sigma ^b{}_{AA'}\nonumber\\
    &=&-\tau _{AA'}{}^b
        +((1/4)+(w/4))(\nabla\cdot\xi )\sigma ^b{}_{AA'}\, ,
\end{eqnarray}    
we find
\begin{eqnarray}
\Lie_\xi\LL &=&
\tau ^{AA'BB'} i2^{-1/2}\left\{
 -{\overline\psi}_{(A'}\nabla _{B)(B'}\psi _{A')}
 +(\nabla _{(B|(B'}{\overline\psi}_{A')})\psi _{|A)} \right.\nonumber\\
&&\qquad\left. -{\overline\psi}_{(A}\nabla _{B)(B'}\psi _{A')}
 +(\nabla _{(B|(B'}{\overline\psi}_A)\psi _{A')} \right\}
   \nonumber\\
 &&+(\nabla\cdot\xi)i2^{-1/2} \left\{
   -((3/4)+(w/4))({\overline\psi}_{A'}\nabla ^{AA'}\psi _A
     -(\nabla ^{AA'}{\overline\psi}_{A'})\psi _A )
     \right.\nonumber\\
&&\qquad\left.     +((1/4)+(w/4))({\overline\psi}^A\nabla _{AA'}\psi ^{A'}
       -(\nabla _{AA'}{\overline\psi}^A)\psi ^{A'})
       \right\} \nonumber\\
  &&+(\nabla\cdot\xi )((1/2)+(w/4))\left(
    \mA (\epsilon ^{AB}\psi _B\psi _A +\epsilon ^{A'B'}{\overline\psi}_{A'}{\overline\psi}_{B'})
    \right.\nonumber\\
&&\qquad\left.     -\mB (\epsilon _{B'A'}\psi ^{B'}\psi ^{A'} +\epsilon _{BA}{\overline\psi}^A{\overline\psi}^B)\right)
    +\text{terms vanishing on-shell} .\qquad
\end{eqnarray}        
After some algebra and use of the field equations, this can be expressed rather simply:
\begin{eqnarray} 
\Lie _\xi\LL &=&\delta g^{AA'BB'} (-i2^{-3/2})
   \left\{
 -{\overline\psi}_{A'}\nabla _{BB'}\psi _{A}
+(\nabla _{BB'}{\overline\psi}_{A'})\psi _{A}\right.\nonumber\\
&&\qquad\left.  -{\overline\psi}_{A}\nabla _{BB'}\psi _{A'}
 +(\nabla _{BB'}{\overline\psi}_A)\psi _{A'} \right\}
 +\text{terms vanishing on-shell}   
  ,\qquad
\end{eqnarray}
where $\delta g^{ab} =-\nabla ^a\xi ^b-\nabla ^b\xi ^a$ is the variation in the inverse metric induced by the diffeomorphism.
We will thus have
\begin{eqnarray}
T_{ab}&=&i2^{-1/2}\sigma _{(a}{}^{AA'}\sigma _{b)}^{BB'}
\left\{
 {\overline\psi}_{A'}\nabla _{BB'}\psi _{A}
 -(\nabla _{BB'}{\overline\psi}_{A'})\psi _{A}
 +{\overline\psi}_{A}\nabla _{BB'}\psi _{A'}
 -(\nabla _{BB'}{\overline\psi}_A)\psi _{A'} \right\}\nonumber\\
 &=&(i/2)\left\{ {\tilde\psi}\gamma _{(a}\nabla _{b)}\psi -(\nabla _{(a}{\tilde\psi}\gamma _{b)})\psi 
 \right\}
 \, .
\end{eqnarray}
This expression is formally the same as the stress--energy for a Dirac particle; however, the field operators depend on the Majorana mass terms as well as the Dirac ones.  It will be locally conserved, that is
\begin{equation}\label{sten}
  \nabla ^aT_{ab}=0\, ,
\end{equation}
precisely by the usual arguments:  the action is invariant and its change is given by
$-\int (\nabla ^a\xi ^b+\nabla ^b\xi ^a) T_{ab}\, d{\rm vol}$; integrating by parts gives equation (\ref{sten}).  One can, of course, verify this equation by a direct calculation, as well.

\section{Discussion}

There are two main lessons of the work here.  The first is that Fermion stress--energies may be derived, at a formal level, from general covariance in a manner patterned on those of Bose fields, and they have the expected forms.  The form of the result is not surprising; the derivation of it by these means is what is of interest.  The question of how quantum fields give gravitational sources is likely a key issue in reconciling quantum theory and gravity, and, as the bulk of familiar matter is Fermionic, we gain a little more evidence for viewing the conventional field-theoretic stress--energy as the likely place to begin this study, as opposed to more exotic proposals like the Kibble--Sciama theory.  (Of course, proponents of such theories often have other reasons for favoring them.)

The second lesson is that there is indeed a calculus of Lie derivatives of spinor fields useful in general relativity.  It is based on the work of Kosmann, and is compatible with the formula of Penrose (for the special case of differentiation along a conformal Killing field); the differences in their formulas are due to differences in their definitions of spinors, but these can be reconciled.  However, the geometric significance of the Lie derivative of a spinor field $\chi ^A$ cannot be considered to reside in $\Lie _\xi \chi ^A$ alone; because of the Lie-theoretically artificial projection involved in forming that quantity, one must consider it together with the Lie derivatives $\Lie _\xi\sigma _a{}^{AA'}$, $\Lie _\xi \epsilon _{AB}$ of the Infeld--van der Waerden symbol and alternating spinor.  Indeed, in the computation of the stress--energy, it was those quantities which gave the only surviving terms.

It is worth noting that the computation of the stress--energy is remarkably lengthy.  Partly that is due to working in curved space--time and allowing both Dirac and Majorana terms, and partly it is due to the projection introduced in computing the Lie derivatives of spinors.  Possibly there is a viewpoint from which this projection is unnecessary and one can get a more direct derivation.

Finally, some comments are in order on the question of just what definition of spinors is used.  As I have pointed out here, the different definitions are all equivalent for any one metric.\footnote{Strictly speaking, for any one metric and choice of equivalence class of spin structure, determined by an element of $H^1(M,\Z _2)$.}  However, the isomorphisms between them depend on the scale of the metric, and so they have different conformal behaviors and lead to different Lie derivatives.  On the other hand, if one is in the end computing a tensorial quantity, that quantity must be independent of the weight $w$ chosen for the definition of the spinor fields.

There are, however, certain purposes for which the choice of weight $w$ could make a difference.  One example would be in the choice of measure for path-integrals over both metrics and Fermions, or the related quantum-statistical relations between gravitational waves and Fermions envisaged in models of the very early Universe.

Another is the behavior of spinors under reflective transformations.  (See ref. \cite{BMDMGK} for closely related material.)
We saw that in general the tensor representing a spinor could have the form $F^{ab}\otimes (X^c)^w$
(where $X^c$ stands for a vector in the null flag modulo vectors in the flagpole direction).  It is apparent from this that factors of $\pm\sqrt{(-1)^{w}}$ are potential contributions to the parity, and these depend on the definition of the spinor adopted.  However, it is {\em not} true that one will simply get an additional such factor.  The reason is that there are other potential contributions to the effects of  reflective motions (and the consequent definitions of spinors), which were not discussed in this paper because the only need was for a treatment of infinitesimal, or small, motions.  If reflective transformations are allowed, there are some further distinctions which need to be considered.  
In particular, because the Penrose--Rindler definition ($w=-1$) of a spinor involves not only knowledge of the null flag but also knowledge of the chirality which is wanted, it is not characterized simply by the tensor product $F^{ab}\otimes (X^c)^w$ if spatial inversions are allowed; in fact its behavior under those motions turns out to be the same as those for the tensor cases $w=0$ and $w=2$.  One should also, for completeness, consider the possibilities of weights associated with timelike or null directions.  

\section*{Data accessibility}

This paper has no data.

\section*{Competing interests}

I have no competing interests.

%

\section*{Acknowledgement}

I thank the referees for useful comments.

\section*{Funding statement}

This work was not supported by external funding.





\begin{thebibliography}{10}
\providecommand{\url}[1]{\texttt{#1}}
\providecommand{\urlprefix}{URL }
\expandafter\ifx\csname urlstyle\endcsname\relax
  \providecommand{\doi}[1]{doi:\discretionary{}{}{}#1}\else
  \providecommand{\doi}{doi:\discretionary{}{}{}\begingroup
  \urlstyle{rm}\Url}\fi
\providecommand{\eprint}[2][]{\url{#2}}

\bibitem{Wald1994}
Wald RM, 1994 \emph{Quantum Field Theory in Curved Spacetime and Black Hole
  Thermodynamics}.
\newblock Chicago: University Press.

\bibitem{ADH1996}
Helfer AD, 1996 The stress--energy operator.
\newblock \emph{Class. Quant. Grav.} \textbf{13}, L129--L134.

\bibitem{Weinberg1995v1}
Weinberg S, 1995 \emph{The quantum theory of fields}, vol.~1.
\newblock Cambridge: Cambridge University Press.

\bibitem{HHKN}
Hehl FW, von~der Heyde P, Kerlick GD, Nester JM, 1976 General relativity with
  spin and torsion: Foundations and prospects.
\newblock \emph{Rev. Mod. Phys.} \textbf{48}, 393--416.
\newblock \doi{10.1103/RevModPhys.48.393}.

\bibitem{Rosenfeld1940}
Rosenfeld L, 1940 Sur le tenseur d'impulsion--\'energie.
\newblock \emph{M\'emoires Acad. Roy. de Belgique}
\textbf{18}, 1--30.

\bibitem{Kibble1961}
Kibble TWB, 1961 Lorentz invariance of the gravitational field.
\newblock \emph{J. Math. Phys.} \textbf{2}, 212.

\bibitem{Sciama1964}
Sciama DW, 1964 The physical structure of general relativity.
\newblock \emph{Rev. Mod. Phys.} \textbf{36}, 463, 1103.

\bibitem{Kosmann1971}
Kosmann Y, 1971 D\'eriv\'ees de {L}ie des spineurs.
\newblock \emph{Annali di Matematica Pura ed Applicata} \textbf{91}, 317--395.
\newblock \doi{10.1007/BF02428822}.

\bibitem{BourguignonGauduchon1992}
Bourguignon JP, Gauduchon P, 1992 Spineurs, op\'erateurs de {D}irac et variations
  de m\'etriques.
\newblock \emph{Communications in Mathematical Physics} \textbf{144}, 581--599.
\newblock \doi{10.1007/BF02099184}.

\bibitem{GodinaMatteucci2003}
Godina M, Matteucci P, 2003 Reductive {G}-structures and {L}ie derivatives.
\newblock \emph{J. Geom. Phys.} \textbf{47}, 66--86.

\bibitem{FatibeneFrancaviglia2003}
Fatibene L, Francaviglia M, 2003 \emph{Natural and Gauge Natural Formalism for
  Classical Field Theories}.
\newblock Dordrecht/Boston/London: Kluwer Academic Publishers.

\bibitem{LRW2014}
Le\~ao RF, Rodrigues Jr WA, Wainer SA, 2014 Concept of {L}ie derivative of spinor
  fields. {A} geometric motivated approach.
\newblock Arxiv:1411.7845 [math-phys].

\bibitem{PR1984}
Penrose R, Rindler W, 1984 \emph{Spinors and space--time, vol. 1: Two--spinor
  calculus and relativistic fields}.
\newblock Cambridge University Press.

\bibitem{PR1986}
Penrose R, Rindler W, 1986 \emph{Spinors and space--time, vol. 2: Spinor and
  twistor methods in space--time geometry}.
\newblock Cambridge University Press.

\bibitem{GodinaMatteucci}
Godina M, Matteucci P, 2005 The {L}ie derivative of spinor fields: theory and
  applications.
\newblock \emph{Int. J. Geom. Meth. Mod. Phys.} \textbf{2}, 159.

\bibitem{Schweber1961}
Schweber SS, 1961 \emph{An Introduction to Relativistic Quantum Field Theory}.
\newblock Row, Peterson and Company.

\bibitem{ChengLi}
Cheng TP, Li LF, 1984 \emph{Gauge theory of elementary particle physics}.
\newblock Oxford: Oxford University Press.

\bibitem{RP1965}
Penrose R, 1965 Zero rest-mass fields including gravitation: asymptotic
  behaviour.
\newblock \emph{Proc. R. Soc. Lond. A.} \textbf{284}, 159--203.

\bibitem{BMDMGK}
Berg M, DeWitt-Morette C, Gwo S, Kramer E, 2001 {The {P}in groups in physics:
  {C}, {P}, and {T}}.
\newblock \emph{Rev. Math. Phys.} \textbf{13}, 953--1034.
\newblock \doi{10.1142/S0129055X01000922}.
\newblock \eprint{math-ph/0012006}.

\end{thebibliography}

\end{document}